\documentclass[aps,prl,twocolumn,showpacs,amsmath,amssymb,floatfix,nofootinbib]{revtex4}

\usepackage{graphicx}
\usepackage{dcolumn}
\usepackage{bm}

\begin{document}
\title{In defense of local textures\\
(and other Higgs gradients)}

\author{Tommy Anderberg}
\email{Tommy.Anderberg@simplicial.net}
\noaffiliation

\date{\today}

\begin{abstract}
Cruz et al. recently showed that the CMB cold spot can be explained
by a GUT-scale texture. But following Turok's argument that gauged
configurations always relax quickly, they posit a global symmetry,
without obvious relation to GUTs. An observation by Nambu
invalidates Turok's argument when the broken symmetry group has
commuting generators. This is demonstrated explicitly in the
standard model of electroweak interactions and holds generally for
intermediate SSB stages in GUTs. The cold spot could therefore be
due to a GUT texture, and electroweak Higgs gradients may evolve
indefinitely.
\end{abstract}

\pacs{11.27.+d,98.80.-k,12.10.-g,11.15.Ex}

\maketitle


\section{Introduction}
Are all gauge interactions unified at some high energy scale?
Running the renormalization group equations of the Minimal
Supersymmetric Standard Model (MSSM) under the assumption that
supersymmetry shows up around $1\thinspace TeV$, the couplings
converge at $1.2\times 10^{16}\thinspace GeV$ \cite{langacker1993},
suggesting the existence of a Grand Unified Theory (GUT) which
undergoes spontaneous symmetry breaking (SSB) at that scale -- alas,
far beyond reach of accelerator experiments. It is therefore
intriguing that the anomalous cold spot found by WMAP in the cosmic
microwave background (CMB) can be explained by the collapse of a
texture (a localized, knot-like field configuration) originating at
a symmetry breaking scale $8.7\times 10^{15}\thinspace GeV$
\cite{cruz2007}.

A window on otherwise inaccessible GUT physics may have opened, but
there is a problem. GUTs are gauge field theories. The texture in
\cite{cruz2007}, on the other hand, is produced by breaking a global
symmetry (``global texture'').

The reason is the following argument, due to Turok \cite{turok1989}:
a texture consists of a scalar multiplet $\Phi(x)$ taking values on
a vacuum manifold $\mathcal{M}$ (the bottom of some symmetric
potential $V(\Phi)$), initially chosen randomly upon SSB in causally
disconnected regions \cite{kibble1976}. It can therefore be written
$\Phi(x) = {\bm U}(x) \Phi_0$, with $\Phi_0$ an arbitrarily chosen
constant on $\mathcal{M}$ and ${\bm U}$ a local symmetry
transformation. If $\Phi$ is gauged, i.e. if it is a Higgs field
\cite{higgs1964}\cite{higgs1966}, the covariant derivative acting on
it is ${\bm D}_\mu(x) =
\partial_\mu + i g {\bm W}_\mu(x)$, with ${\bm W}_\mu$ denoting
the gauge fields and $g$ the coupling constant. ${\bm W}_\mu$ can
now ``fall'' to ${\bm W}_\mu = (i/g) \left(\partial_\mu {\bm
U}\right){\bm U}^{-1}$ at every point in space, making ${\bm D}_\mu
\Phi = 0$. The gradient energy of $\Phi$ is then zero. ${\bm W}_\mu$
is pure gauge, so its energy also vanishes and the configuration
stops evolving.

Naively, this relaxation process should play out on the time scale
typical of the gauge interaction, i.e. in a microphysical time,
whereas the texture in \cite{cruz2007} created the CMB cold spot by
collapsing at redshift $z \sim 6$, a billion years after the big
bang. So at first sight it can only be a global texture, and it is
not at all clear how it might fit in a GUT scheme.

\section{Two caveats}

A first caveat to the naive guesstimate above was pointed out in
\cite{anderberg2005}: even topologically trivial $\Phi$
configurations constrained to $\mathcal{M}$ carry conserved
quantities (energy, momentum, gauge currents, all in derivative
terms) which must go elsewhere, i.e. to fermions, upon relaxation.
Fermions can only be produced effectively while the energy and
charges within the Compton volume of a fermion pair (e.g. electron +
neutrino for electroweak interactions) are $\geq$ the total mass and
charges of such a pair (on shell). Once $\Phi$ gradients fall below
this threshold, dissipation to fermions becomes exponentially
suppressed. The evolution of $\Phi$ and ${\bm W}_\mu$ then becomes a
Hamiltonian flow, but need not stop.

A second, subtler caveat follows from Nambu's ``generalized Meissner
effect'' \cite{nambu1977}: the condition ${\bm D}_\mu\Phi = 0$ does
not guarantee vanishing energy if a symmetry generator has a zero
eigenvalue in the representation of $\Phi$, i.e. if a subset ${\bm
A}_\mu$ of ${\bm W}_\mu$ remains massless after SSB. To see this,
apply the transformation law $ {\bm W}_\mu \to {\bm W}'_\mu = {\bm
U}{\bm W}_\mu{\bm U}^{-1} + (i/g)\left(\partial_\mu {\bm
U}\right){\bm U}^{-1}$ to a constant ${\bm W}_\mu = {\bm A}_\mu$.
Rather than the unique ${\bm W}_\mu$ vacuum implicitly assumed by
the naive argument, there is now a manifold of degenerate vacua
${\bm W}_\mu = {\bm U}{\bm A}_\mu{\bm U}^{-1} + (i/g)
\left(\partial_\mu {\bm U}\right){\bm U}^{-1}$. If each point in
space picks ${\bm W}_\mu$ independently, there will be (generalized)
electric and magnetic fields with finite energy density. Relaxation
to vacuum can therefore not proceed independently at each point: the
relaxation rate is limited by causality. This happens when the group
is not simple or the representation's rank is $\geq 2$, i.e. for any
realistic gauge field theory, since linear combinations of commuting
generators then exist which annihilate $\Phi$.

\section{Non-Linear Sigma Model}

The global texture in \cite{cruz2007} is modeled by an $O(4)$
non-linear sigma model (NLSM), the classical theory of a real
4-component $\Phi$ constrained to take values on a 3-sphere.
The classical approximation can be motivated by noting that in the
low energy/long wavelength limit, the quantum effective action is
dominated by stationary points of the action; SSB provides a simple
example of background field quantization, with quantized short
wavelength perturbations, i.e. particles, propagating over a
semiclassical background of long wavelength modes
\cite{dewitt1967}\cite{grigoriev1988}. The NLSM approximation holds
because excitations in the non-flat direction of $V(\Phi)$ are
suppressed by powers of interaction energy $E_I$ over symmetry
breaking scale $E_{SB}$. In practice the Lagrangian is split in two:
a low energy semiclassical part for long wavelength modes with large
occupation number (implying mass $\ll E_I$) and a high energy UV
completion for particles.

All this remains true when $\Phi$ is a Higgs field. For $E_I \ll$
Higgs mass, a Higgsed gauge theory reduces to a gauged NLSM (GNLSM),
which can be viewed as the leading, model-independent term of an
effective Lagrangian with model-dependent higher order terms in
$E_I/E_{SB}$ \cite{appelquist1975}. Since it does not include radial
excitations, this GNLSM remains valid (and classically exact) even
if the Higgs field is only effective and no Higgs particle exists,
as in technicolor models. In the electroweak example, since the
minimal model is experimentally confirmed up to $E_I \sim
100\thinspace GeV$ \cite{yao2006}, whatever actually causes
electroweak symmetry breaking must reduce to the same GNLSM for $E_I
\ll 100\thinspace GeV$, or equivalently for distances $\gg
10^{-18}\thinspace m$, with any corrections from unknown sectors
suppressed by powers of $\sim E_I/(100\thinspace GeV)$
\cite{appelquist1980}\cite{longhitano1980}\cite{longhitano1981}.

The electroweak GNLSM is conventionally written in ``polar'' field
coordinates $\theta_a$, so that
\begin{eqnarray}\label{su2U}
{\bm U} = \exp\left(i \theta_a \tau_a / 2\right) = \cos(\theta/2) +
i\thinspace \frac{\theta_a\tau_a}{\theta}\sin(\theta/2)
\end{eqnarray}
with $\tau_a =$ Pauli matrices, $\theta =
\sqrt{\left(\theta_1\right)^2 + \left(\theta_2\right)^2 +
\left(\theta_3\right)^2}$ and
\begin{eqnarray}\label{phiThetaEquivalence}
\left[
\begin{array}{lr}
\phi^{0\dag} & \phi^+\\
-\phi^{+\dag} & \phi^0
\end{array}
\right] = \frac{\nu}{\sqrt{2}} {\bm U}
\end{eqnarray}
where $\nu \simeq 246.3\thinspace GeV$ is the symmetry breaking
parameter. The covariant derivative acting on ${\bm U}$ is
\begin{eqnarray}\label{DU}
{\bm D}_\mu = \partial_\mu + i \frac{g_W}{2} W^a_\mu \tau^a - i
\frac{g_B}{2} B_\mu \tau^3
\end{eqnarray}
and the full Lagrangian is
\begin{eqnarray}\label{polarLagrangian}
{\cal L}_{EW} &=& -\frac{1}{4}B_{\mu\nu}B^{\mu\nu} - \frac{1}{4}W^a_{\mu\nu}W^{a\mu\nu} \nonumber \\
&&+ \frac{\nu^2}{4} {\rm Tr}\left[\left({\bm D}_\mu {\bm
U}\right)^\dag \left({\bm D}^\mu {\bm U}\right)\right]
\end{eqnarray}
${\cal L}_{EW}$ has been used as is to compute the one-loop thermal
effective action for an electroweak plasma at $E_I$ between Higgs
and weak gauge boson masses $m_W$ and $m_Z$ \cite{manuel1998}, but
at low $z$ we are interested in the low energy limit $E_I \ll m_W <
m_Z$, where the massive gauge bosons can not be excited either and
the semiclassical background consists of massless modes only.
Remembering Nambu's lesson, we therefore look for a linear
combination of gauge fields with zero effective mass, i.e. a
generalized photon.

In the basis $\left[B_\mu, W^1_\mu, W^2_\mu, W^3_\mu\right]$, the
${\cal L}_{EW}$ terms quadratic in $B_\mu$ and $W^a_\mu$ give rise
to the mass matrix
\begin{eqnarray}\label{gaugeMassMatrix}
\frac{\nu^2}{2}
\left[
\begin{array}{cccc}
g_B^2 & g_B g_W \Theta_1 & g_B g_W \Theta_2 & -g_B g_W \Theta_3 \\
g_B g_W \Theta_1 & g_W^2 & 0 & 0 \\
g_B g_W \Theta_2 & 0 & g_W^2 & 0 \\
-g_B g_W \Theta_3 & 0 & 0 & g_W^2
\end{array}
\right]
\end{eqnarray}
where we have introduced the convenient auxiliary quantities
\begin{eqnarray}
\Theta_1 &=& \left[\theta_1\theta_3(\cos(\theta)-1)+\theta\theta_2\sin(\theta)\right]/\theta^2 \label{Theta1} \\
\Theta_2 &=& \left[{\theta_2\theta_3}(\cos(\theta)-1)-{\theta\theta_1}\sin(\theta)\right]/\theta^2 \label{Theta2} \\
\Theta_3 &=& \left[(\theta_1^2+\theta_2^2)\cos(\theta) + \theta_3^2\right]/\theta^2 \label{Theta3}
\end{eqnarray}
satisfying $\left(\Theta_1\right)^2 + \left(\Theta_2\right)^2 +
\left(\Theta_3\right)^2 = 1$. The eigenvalues of Eq.
\eqref{gaugeMassMatrix} are the tree level masses squared of photon,
$W^{\pm}$ and $Z^0$. The two degenerate eigenstates can be
orthogonalized to obtain
\begin{eqnarray}
A_\mu           &\propto& \left[\frac{g_W}{g_B \Theta_3}, -\frac{\Theta_1}{\Theta_3}, -\frac{\Theta_2}{\Theta_3}, 1\right] \label{eigenA} \\
\acute{W}^1_\mu &\propto& \left[0, -\frac{\Theta_2}{\Theta_1}, 1, 0\right] \label{eigenW1} \\
\acute{W}^2_\mu &\propto& \left[0, \frac{\Theta_1 \Theta_3}{\Theta_1^2 + \Theta_2^2}, \frac{\Theta_2 \Theta_3}{\Theta_1^2 + \Theta_2^2}, 1\right] \label{eigenW2} \\
Z_\mu           &\propto& \left[-\frac{g_B}{g_W \Theta_3}, -\frac{\Theta_1}{\Theta_3}, -\frac{\Theta_2}{\Theta_3}, 1\right] \label{eigenZ}
\end{eqnarray}
Inverting Eqs. \eqref{eigenA}-\eqref{eigenZ} and setting
$\acute{W}^1_\mu = \acute{W}^2_\mu = Z_\mu = 0$ to account for the
decay of all massive modes yields
\begin{eqnarray}
B_\mu &=& A_\mu \cos(\theta_W) \label{bFromA} \\
W^1_\mu &=& -A_\mu \Theta_1 \sin(\theta_W) \label{w1FromA} \\
W^2_\mu &=& -A_\mu \Theta_2 \sin(\theta_W) \label{w2FromA} \\
W^3_\mu &=&  A_\mu \Theta_3 \sin(\theta_W) \label{w3FromA}
\end{eqnarray}
where $\theta_W$ is the Weinberg angle, $\sin(\theta_W) = g_B /
\sqrt{g_B^2 + g_W^2} \simeq \sqrt{0.2216}$. Substituting Eqs.
\eqref{bFromA}-\eqref{w3FromA} into ${\cal L}_{EW}$ then yields our
final, low energy effective Lagrangian for the electroweak boson
sector
\begin{eqnarray}\label{masterLagrangian}
{\cal L} &=& \frac{\nu^2}{8}\left[ \partial_\mu \theta \partial^\mu \theta
                                  + \frac{4 \sin^2(\theta/2)}{\theta^2}
                                    \left( \frac{\vec{\theta}}{\theta} \times \partial_\mu \vec{\theta}\right)^2 \right] \nonumber \\
&&- \frac{1}{4} \left( \partial_\mu A_\nu - \partial_\nu A_\mu \right)^2 \nonumber \\
&&- \frac{\sin^2(\theta_W)}{4} \left( A_\mu \partial_\nu \Theta_a - A_\nu \partial_\mu \Theta_a \right)^2
\end{eqnarray}
with $\vec{\theta} = [\theta_1, \theta_2, \theta_3]$, $\theta = |\vec{\theta}|$.

The first row in Eq. \eqref{masterLagrangian} is just the plain
$O(4)$ NLSM in polar field coordinates, the second row is the
Maxwell Lagrangian, the third row couples them $\propto
\sin^2(\theta_W)$, acting as an effective photon mass term when
$\vec{\theta}$ is not constant. Analogous Lagrangians (with larger
symmetries and more fields) can be expected to describe the long
wavelength modes of GUTs at intermediate symmetry breaking stages.

\section{Gauge fixing}

Eq. \eqref{masterLagrangian} does not include gauge fixing terms; a
physical gauge, i.e. a gauge which does not introduce fictitious
fields, is therefore implied. Most convenient (and popular) is the
time-axial gauge $A_0 = 0$
\cite{nielsen1973}\cite{dashen1974}\cite{klinkhamer1984}\cite{ambjorn1991}\cite{galtsov1991}\cite{graham2007}
(note that in quantum theory, the unitary gauge is the singular
limit of 't Hooft's $R_\xi$ gauges, 
which require the introduction of ghosts). The electric and magnetic
fields are then $\vec{E} = -\partial_0\vec{A}$, $\vec{B} = \nabla
\times \vec{A}$, and the energy density is a manifestly non-negative
sum of quadratic forms,
\begin{eqnarray}\label{rho}
\rho &=& \frac{1}{2}\left(\vec{E}^2 + \vec{B}^2\right) \nonumber \\
&&+ \frac{1}{2} \partial_0\vec{\theta}^T \left(\nu^2 {\bm G} + \vec{A}^2 {\bm H} \right) \partial_0\vec{\theta} \nonumber \\
&&+ \frac{\nu^2}{2} \partial_m \vec{\theta}^T {\bm G}
\partial_m \vec{\theta} \nonumber \\
&&+ \frac{1}{2} \left(\varepsilon_{jkl} A_k \partial_l
\vec{\theta}\right)^T {\bm H} \left(\varepsilon_{jmn} A_m \partial_n
\vec{\theta}\right)
\end{eqnarray}
where ${\bm G}$, with components
\begin{eqnarray}\label{G}
G_{ab} = \left(\frac{\delta_{ad}\delta_{be}}{2} + \frac{1 -
\cos{\theta}}{\theta^2} \varepsilon_{cda}\varepsilon_{ceb} \right)
\frac{\theta_d\theta_e}{\theta^2}
\end{eqnarray}
is the 3-sphere metric, with eigenvalues $1/4$ and (doubly
degenerate) $0 \leq (1-\cos(\theta))/(2\theta^2) \leq 1/4$, while
${\bm H}$, with components
\begin{eqnarray}\label{H}
H_{ab} = \sin^2(\theta_W) \frac{\partial\Theta_c}{\partial\theta_a}
\frac{\partial\Theta_c}{\partial\theta_b}
\end{eqnarray}
also has no negative eigenvalues (but zeros along all axes of
$\vec{\theta}$ space). By the spectral theorem, $\vec{A} \neq 0$ can
therefore only increase $\rho$ for a given $\vec{\theta}$.

Varying ${\cal L}$ in $A_0$ yields the Gauss constraint
\begin{eqnarray}\label{gauss}
\partial_0\partial_m A_m = \partial_0 \vec{\theta}^T {\bm H} A_m \partial_m
\vec{\theta}
\end{eqnarray}
(a plane in $\vec{A}$ space) which completely fixes the gauge,
leaving $\vec{A}$ with only two independent degrees of freedom. As
in any gauge other than unitary, the Goldstone modes remain. Keep in
mind that we are working with non-perturbative, background
\emph{fields}; perturbations on this background (described by the UV
completion not shown here) damp out within a Compton wavelength of
the massive gauge bosons, so there is no plague of massless
Goldstone \emph{particles} propagating over macroscopic distances
(see e.g. \cite{dams2004} for an explicit demonstration in the
space-axial gauge). Those are confined, either to the background or
to individual massive gauge bosons.

\section{Discussion}
Since $\vec{A} \neq 0$ can only increase $\rho$ for a given
$\vec{\theta}$, it is tempting to set $\vec{A}= 0$ and coopt the
wealth of existing work on the plain NLSM from hadron physics
\cite{cornwall1979}\cite{blaizot1992}\cite{huang1996}\cite{ioannidou2004},
cosmology \cite{phillips1995}, gravity
\cite{duan1983}\cite{liebling2000}\cite{liebling2002}\cite{ward2002}
and general mathematical physics
\cite{fuller1954}\cite{corlette1999}\cite{bizon1999}\cite{bizon2000}\cite{ward2003}\cite{wereszczynski2005}.
From this work, we know families of solutions on various metrics
(and that any solution of the massless Klein-Gordon equation on the
target spacetime \cite{feng2001} can be used to generate solutions
along geodesics of $\mathcal{M}$ \cite{duan1983}); that
singularities never occur for suitably small initial data
\cite{liebling2000}, making total decay to fermions unlikely (pair
production makes large data small, but shuts down below a
threshold); and that collapse can be prevented by rapid metric
expansion \cite{ward2002}. From the point of view of
\cite{anderberg2005}, the existence of harmonic maps with polyhedral
symmetry \cite{ioannidou2004}\cite{kleihaus2004} is particularly
interesting.

But we also know that electroweak texture collapse can be prevented
by gauge interactions \cite{turok1990}, and as Nambu taught us,
there may be local minima with $\vec{A}\neq 0$ (e.g. cosmic magnetic
fields coupled to $\vec{\theta}$ vortices). For GUTs, with more
massless gauge fields at intermediate SSB stages, the space of
possible solutions is also larger (but there may also be NLSM
subgroups which do not couple directly to the photon).

Even the lowly standard model may have more surprises in store. To
start with, the $\sin^2(\theta_W)$ term in Eq.
\eqref{masterLagrangian} couples $\vec{\theta}$ to a heat bath of
CMB photons. By inspection of the equations of motion, any NLSM
solution satisfying $\partial_\mu\partial^\mu\theta_a =
\partial_\mu\theta_a\partial^\mu\theta_b = 0$ (e.g. a plane wave)
is also a (stochastic) GNLSM solution for constant $\langle A_m
\rangle = 0$, $\langle A_m A_n \rangle = A^2 \delta_{mn}$, but what
about the general case? Does radiation pressure significantly affect
the dynamics, e.g. driving transition layers between regions of
constant $\vec{\theta}$ (domain boundaries) to $\theta = 0, 2\pi$,
where $\vec{A}$ decouples?

When fermions are included, there is also a neutrino heat bath to
consider, along with the real jokers: the quark condensates $\langle
\bar{q}q \rangle$ which form at the chiral phase transition of QCD
\cite{ioffe2006}, with their own NLSM coupled to the GNLSM through
${\bm D}_\mu$ and Yukawa terms. Intriguingly, it is evident by
inspection of the latter that the manifest $\mathbb{Z}_2$ symmetry
of Eq. \eqref{masterLagrangian} under $\vec{\theta} \rightarrow
-\vec{\theta}$ is broken by $\langle \bar{q}q \rangle \neq 0$. In
the two-flavor case, with condensates $(\sigma, \pi_a \tau_a)$, it
is also easily verified that $\theta = 0$ (by definition, our
vacuum) can not be a minimum of the QCD-induced potential unless
$\pi_a = 0$, leaving only $\sigma \neq 0$. Since the phases of quark
mass terms and condensates are related to the strong CP-violation
parameter $\theta_{QCD}$ through the axial anomaly
\cite{huang1992}\cite{huang1992b}, it is fair to wonder whether a
solution of the long-standing strong CP problem may be lurking in
this interplay between $\langle \bar{q}q \rangle$ and
$\vec{\theta}$. Could the latter stand in for the stubbornly
undetected and theoretically problematic \cite{kamionkowski1992}
global axion \cite{peccei1977}?

Finally, gravity may be important both locally, as a stabilizer
counteracting dispersion
\cite{volkov1999}\cite{ioannidou2006}\cite{verbin2007}, and
cosmologically. The pressure to $\rho$ ratio of an isotropic
``fluid'' of scalar field configurations goes from $1/3$ for
relativistic wave fronts to $-1/3$ in the static limit. A flat FRW
cosmology dominated by such configurations could therefore evolve
naturally from radiative (scale factor $a(t) \propto \sqrt{t}$)
through ``dust matter'' ($a(t) \propto t^{2/3}$) to linear ($a(t)
\propto t$). Incidentally, a linearly coasting cosmology
\cite{kolb1989} is known to fit observation well
\cite{pimentel1999}\cite{dev2002}\cite{gehlaut2003} while avoiding
some problems of the concordance model
\cite{sethi2005}\cite{jain2006}. A major objection might be that the
photon conversion mechanism of \cite{anderberg2005} would then
result in excessive SNe Ia dimming.


\begin{acknowledgments}
I am grateful to Thomas Schaefer, who vehemently disagrees with this
paper, for calling my attention to a boneheaded error in the first
version.
\end{acknowledgments}

\bibliography{basename of .bib file}

\end{document}